\documentclass[conference]{IEEEtran}
\IEEEoverridecommandlockouts
\usepackage{cite}
\usepackage{amsmath,amssymb,amsfonts}
\usepackage{algorithmic}
\usepackage{graphicx}
\usepackage{textcomp}
\usepackage{xcolor}
\usepackage{subcaption}
\usepackage{todonotes}
\usepackage{hyperref}
\hypersetup{
    colorlinks=true,
    linkcolor=blue,
    linkbordercolor=blue,
    filecolor=magenta,      
    urlcolor=cyan,
}
\usepackage{multicol}
\usepackage[all]{nowidow}
\usepackage{algorithm}

\newtheorem{defi}{Definition}

\def\BibTeX{{\rm B\kern-.05em{\sc i\kern-.025em b}\kern-.08em
    T\kern-.1667em\lower.7ex\hbox{E}\kern-.125emX}}
\begin{document}

\title{On the Importance of demand Consolidation in Mobility on Demand
}

\author{
\IEEEauthorblockN{Andrea Araldo}
\IEEEauthorblockA{\textit{T\'el\'ecom SudParis} \\
\textit{Institut Polytechnique de Paris} \\
Evry, France \\
\small{andrea.araldo@telecom-sudparis.eu}}
\and
\IEEEauthorblockN{Andrea Di Maria}
\IEEEauthorblockA{\textit{Universit\`a di Catania} \\
Catania, Italy \\
\small{a.dimaria@studium.unict.it}}
\and
\IEEEauthorblockN{Antonella Di Stefano}
\IEEEauthorblockA{\textit{Universit\`a di Catania}\\
Catania, Italy \\
\small{ad@dieei.unict.it}}
\and
\IEEEauthorblockN{Giovanni Morana}
\IEEEauthorblockA{\textit{Aucta Cognitio R\&D Labs} \\
Catania, Italy \\
\small{gmorana@auctacognitio.net}}
}

\maketitle

\begin{abstract}
Mobility on Demand (MoD) services, like Uber and Lyft, are revolutionizing the way people move in cities around the world and are often considered a convenient alternative to public transit, since they offer higher Quality of Service (QoS - less waiting time, door-to-door service) at a cheap price. In the next decades, these advantages are expected to be further amplified by Automated MoD (AMoD), in which drivers will be replaced by automated vehicles, with a big gain in terms of cost-efficiency. MoD is usually intended as a door-to-door service. However, there has been recent interest toward consolidating, e.g., aggregating, the travel demand by limiting the number of admitted stop locations. This implies users have to walk from/to their intended origin/destination. 

The contribution of this paper is a systematic study the impact of consolidation on the operator cost and on user QoS. We introduce a MoD system where pick-ups and drop-offs can only occur in a limited subset of admitted stop locations. The density of such locations is a system parameter: the less the density, the more the user demand is consolidated. 
We show that, by decreasing stop density, we can increase system capacity (number of passengers we are able to serve). On the contrary, increasing it, we can improve QoS. The system is tested in AMoDSim, an open-source simulator. The code to reproduce the results presented here is available on-line.

 This work is a first step toward flexible mobility services that are able to autonomously re-configure themselves, favoring capacity or QoS, depending on the amount of travel demand coming from users. In other words, the services we envisage in this work shift their operational mode to any intermediate point in the range from a taxi-like door-to-door service to a bus-like service, with few served stops and more passengers on-board.
\end{abstract}


%%%%%%%%%%%%%%%%%%%%%%%%%%%%%%%%%%%%%%%
%%%%%%%%%%%%%% INTRO %%%%%%%%%%%%%%%%%%
%%%%%%%%%%%%%%%%%%%%%%%%%%%%%%%%%%%%%%%
\section{Introduction}

\begin{figure}[h]
  \centering
	\includegraphics[width=0.6\linewidth]{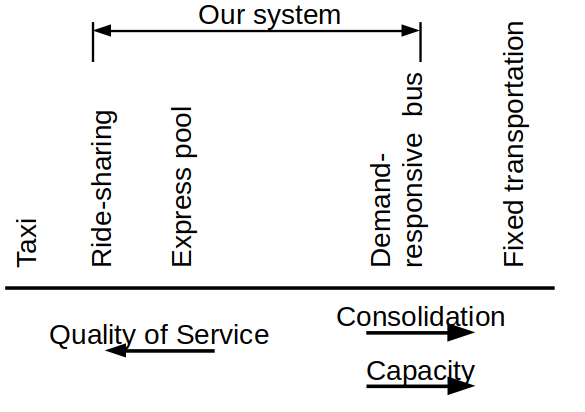}
	\caption{The landscape of transportation. }
    \label{fig:landscape}
\end{figure}

The landscape of transportation services (Fig.~\ref{fig:landscape}) has become in recent years much richer than some decades ago. We identify in it two poles: on the one side \emph{Mobility on Demand} (MoD) and on the opposite side \emph{fixed transportation}. These two poles differ in the way they adapt to the demand. In MoD vehicles do not have pre-determined fixed routes, which are instead constructed based on the trip request. The opposite holds for fixed transportation (bus, subway, trains). Historically, MoD has been represented by taxis, serving mostly one passenger at a time, which implies (i) high cost of operation and thus (ii) high price for the user and  (iii) low \emph{capacity}, i.e., number of passengers that can be served per unit of time. On the other extreme, fixed transportation aims to \emph{consolidate} demand in time and space: users need to walk to/from a limited set of stops (spacial consolidation) that vehicles serve in a limited set of time instants (time consolidation). Consolidation generally achieves high capacity, low cost of operation and thus reasonable levels of price, compatible with the missions of a public service. However, this is achieved in exchange for a reduction in QoS.

The distance between taxis and fixed transportation has been recently filled by ride sharing services, like Uber and Lyft, which can afford to propose cheaper prices, thanks to the ability of consolidating trips: by merging several user trips in the same optimized route, the cost of operating one vehicle is split among different users. This effect is expected to be further amplified if the number of stops is limited, since the consolidation of demand around them allows to construct more efficient routes. This motivated the launch of the experimental Uber Express Pool in November 2017~\cite{Lo2018}, which ``offers the cheapest fares yet in exchange for a little walking'' (\cite{Hawkins2018}).
However, the capacity of ride sharing systems is still far from transit. This motivates a rich research on responsive buses~\cite{Quadrifoglio2009, Guo2018,Koh2018, Chandra2013}, aiming to combine the flexibility of MoD with high capacity. A question has been raised in many research studies of whether to deploy flexible or fixed route buses depending on the amount of demand. Articles as~\cite{Quadrifoglio2009} compare the performance of the two extreme bus services, i.e., fixed vs. on-demand, and found that after a certain request rate threshold the former is superior. We interpret this result under the lenses of consolidation: when high capacity is needed, fixed bus consolidates the demand better than flexible routes.

What commented so far shows that consolidation is becoming a fundamental aspect in future flexible transportation systems.
The contribution of this paper is a systematic study of the role of consolidation in MoD. To this aim, we introduce a MoD system that changes behavior depending on the value of admitted stop density: instead of admitting pick-ups and drop-offs everywhere, we limit them to occur only at certain predefined locations. Their density is the parameter determining consolidation (the more the stop density the more the amount of consolidation). We use this parameter as a knob to explore the trade-off between QoS (travel times suffered by users) and high capacity. We show that by just leveraging this knob, with no change in the vehicle dispatch logic, the system effectively changes behavior to shift its configuration toward the left or the right poles of the transportation landscape.

We simulate the system in AMoDSim~\cite{DiMaria2018} and made all the scripts to reproduce the scenarios available in the repository. We find that, under the assumptions considered here, MoD can effectively offer higher capacity if more consolidation is adopted in change of a reasonable loss in QoS. This holds true with high levels of demand, while consolidation is detrimental otherwise.

%%%%%%%%%%%%%%%%%%%%%%%%%%%%%%%%%%%%%%%%%%%%
%%%%%%%%%%%%% RELATED WORK %%%%%%%%%%%%%%%%%
%%%%%%%%%%%%%%%%%%%%%%%%%%%%%%%%%%%%%%%%%%%%
\section{Related Work}\label{sec:related}
In this section we will first clarify the terminology around MoD and then discuss the work in which the important of consolidation emerges.

The transportation systems we study are given different names in the literature: Mobility on Demand~\cite{Basu2018b}, Dial-a-Ride (see~\cite{Berbeglia2010a} and \S 1.3.1 of \cite{Nunez2013} ), Demand-Responsive~\cite{Winter2016a}, Ride Sharing~\cite{Alonso-Mora2017}, Taxi-Sharing~\cite{Watel2018}, Ride-Hailing~\cite{Bosch2018}.
The common figure is that vehicles do not have pre-determined routes, which are instead constructed on the fly based on requests received from users, either via a smartphone app or, in early days experiments, via telephone calls~\cite{Stein1978}.

The research community has not yet aligned on the the same terminology. Moreover, different nuances exist between the different terms.
Dial-a-ride and Demand-responsive transportation historically denote buses with dynamically determined routes, whose experiments started in the 70s, but which are rarely adopted in reality. The terms Ride-Sharing or Taxi-Sharing are now employed to denote systems like Uber and Lyft operated with taxi-like vehicles with low capacity. The system we studied here spans the range between ride-sharing taxis and demand-responsive buses: by limiting the number of stops we move from the former to the latter, increasing system capacity.

MoD systems need a driver for each vehicle, who works directly or indirectly for the operator and drives only to satisfy the mobility needs of the passengers. There exist other systems, known as \emph{carpooling}, in which the driver owns the car, drives toward her intended destination and can make detours to pick-up and drop-off other passengers. Differently from MoD, in carpooling drivers do not work for the service operator and they primarily satisfy their own mobility needs. Carpooling systems are out of the scope of our paper. We warn the reader that some authors~\cite{Stiglic2015} call them ``ride-sharing'', which may generate confusion.

From the Operations Research viewpoint, the problem that MoD systems need to solve is a Dynamic Pickup and Delivery Problem~\cite{Berbeglia2010}, which is a class of Dynamic Vehicle Routing problems.  The term ``dynamic'' means that the requests are not known in advance but enter the system while it is working and already serving other requests. These kind of problems require thus to continuously modify vehicle routes. It is usually assumed that users are picked-up and dropped-off at their origin and destination, respectively. In this paper we are instead interested in dealing with another decision variable for the operator: the amount of \emph{admissible stop locations} in which we restrict pick-ups and drop-offs. The impact of consolidating the demand by restricting stop locations lacks a systematic study in the literature. This work is a contribution toward filling this gap. 
We will now briefly survey some work in which the importance of consolidation emerges, although not explicitly mentioned and not systematically studied.

Quadrifoglio et Al.~\cite{Quadrifoglio2009} shows the existence of a \emph{critical demand density}, in terms of requests per hour per Km2, over which a demand-responsive service becomes inefficient and it is better to deploy a fixed-route bus. We read in this main finding of the paper the emergence of the benefit of consolidating demand around the few stops when high demand density (aka high capacity) has to be served. Their work is based on an idealized scenario based on the continuous approximation approach, where the studied area is a continuous space (in lieu of a road network) and demand is a continuous density function over that space. 
Differently from their work (i) we can vary the degree of consolidation by varying the admitted stop location density, (ii) we preserve the flexibility of vehicle routes across those stops, even when their density is low, (iii) the area under study is a network and the requests are generated from a discrete stochastic process, which is more realistic, (iv) many vehicles are coordinated to serve the same area, while in~\cite{Quadrifoglio2009} only one vehicle is assumed. 

Stiglic et Al.~\cite{Stiglic2015} study a carpooling system and tackle the problem of fixing convenient meeting points to make drivers and passengers meet. We interpret meeting points as a way to consolidate demand. Differently from~\cite{Stiglic2015}, (i) we study a MoD system, so we do not have the constraints of a pre-defined pair of origin-destination of the driver; (ii) in \cite{Stiglic2015} only one pick-up and one drop-off point can be added to the pre-defined route of the driver, while the routes in our systems can have an arbitrary number of stop-points; (iii) we assume we do not know future requests, while in \cite{Stiglic2015} trips are announced in advance and the matching is performed off-line.

The benefits of consolidation also emerge in fixed-route transportation systems~\cite{EI-Geneidy2007,Li2018a}.
The former found that the consolidation program implemented in Portland, Oregon, consisting in increasing stop spacing, improved bus running times by 6\% and at the same time had no relevant impact on user activity. The latter constructs fixed routes based on hubs that aggregate demand.

None of the previously mentioned research studies the inter-dependence between the system performance, the demand density and the density of stop-locations, which is the focus of our paper.

%%%%%%%%%%%%%%%%%%%%%%%%%%%%%%%%%%%%%%
%%%%%%%%%%%%%%% SYS MODEL %%%%%%%%%%%%
%%%%%%%%%%%%%%%%%%%%%%%%%%%%%%%%%%%%%%
\section{System model}
\label{sec:model}

The system model considered in this paper is an extension of the one presented in our previous work~\cite{DiMaria2018}. The system is composed by a fleet of \emph{vehicles} and a \emph{coordinator} managing them. Users send trip requests (\S~\ref{sec:requests}) to the coordinator which matches them to the available vehicles and modifies the routes (\S~\ref{sec:schedules}), based on a dispatching algorithm (\S~\ref{sec:algorithm}). Note that, however, the results of this paper would still be valid under a different computation model, e.g., also if the user-to-vehicle matching  were not performed by a centralized coordinator but were instead computed in a fully distributed manner, which is outside the scope of this paper. The system is configured to serve a set of \emph{admitted stop locations} $\mathcal{Q}$. For any given location $z$, we denote with $\phi(z)\in\mathcal{Q}$ the closest of the admitted locations. Given two location $q,q'$, we denote with $\tau_w(q,q')$ and $\tau_v(q,q')$ the time needed to travel the shortest path between $q$ and $q'$ by walking and by car, respectively. For simplicity, we assume the speed of walking and of that of cars is fixed and constant.

\subsection{Stop points and Trip requests}
\label{sec:requests}
The elementary unit of information handled by the dispatching algorithm is the \emph{stop point}. It is a tuple $sp_1 = (q,t,\Delta t,a)$, where $q\in \mathcal{Q}$ is a \emph{location} amongst the admitted, $t$ is a \emph{preferred time}, $\Delta t$ is the \emph{maximum extra time} and $a$ is a binary variable specifying if the stop point is a pick-up or a drop-off. A stop point denotes that a user wishes to be picked-up or dropped-off at location $q$ in the inteval $[t,t+\Delta t[$. The dispatching algorithm that we will discuss later constructs vehicle routes as a sequence of stops, each stop corresponding to some stop points, and ensures the time constraints are not violated.

A \emph{trip} is described by an origin location $o$, a destination location $d$, a time $t$ denoting the time at which the request has been generated and a time constraint $\Delta t$. A user must go to the admitted stop location $\phi(o)$ closest to the origin, send a request and wait for a vehicle. She then takes a ride to the admitted location $\phi(d)$ closest to the intended destination, toward which she walks. Therefore, a trip corresponds to a pair of stop points: a pick-up stop point $sp^a = \left( \phi(o),t_1,\Delta t, \text{pickup} \right)$ and a drop-off stop point $sp^b = \left(\phi(d),t_2,\Delta t, \text{dropoff}\right)$. Here, $t_1$ is the time instant in which the user arrives at the stop location $\phi(o)$. Let us denote with $t$ the time at which the user ``appears'', i.e., she starts to walk from her origin, and with $\tau_w \left({o,\phi(o) }\right)$ the walking time from $o$ to $\phi(o)$. We have $t_1 = t+\tau_w(o,\phi(o))$. Denote with $\tau_v \left(\phi(d), d \right)$ the minimum time to go from $\phi(d)$ to $d$ by car, we have $t_2=t_1+\tau_v \left(\phi(d), d \right)$. We  call $\tau_v \left(o, \phi(o) \right)$ and $\tau_v \left(\phi(d), d \right)$ \emph{ingress} and \emph{egress} time, respectively.

To clarify the meaning of the time constraint $\Delta t$, it must be observed that the dispatching algorithm will try to place $sp^a$ and $sp^b$ into the vehicle routes (we will explain this in \S~\ref{sec:algorithm}). The specification of stop-point $sp^a$ will force the system to pick-up the passenger from $\phi(o)$ in the interval $[t_1,t_1+\Delta t[$. Similarly, the specification of stop-point $sp^b$ forces the dispatching algorithm to drop-off the user at $\phi(d)$ in the interval $[t_2, t_2+\Delta t[$. If it is impossible to satisfy any of the two conditions, the request is discarded.

\subsection{Vehicle schedules}
\label{sec:schedules}
At any time instant, each vehicle $v$ is associated with a \emph{schedule} $S_v = [sp_1, sp_2,\dots]$, which is the sequence of stop points the vehicle plans to serve. This sequence is associated with a time sequence $T_v=[\hat t_1, \hat t_2, \dots]$ denoting that stop point $sp_i=(q_i, t_i, \Delta_i, a_i)$ will be served at time $\hat t_i$. In order to satisfy the time constraints, the following condition must be satisfied
\begin{flalign}
    \label{eq:constr1}
    t_i \le \hat t_i < t_i + \Delta t_i.
\end{flalign}

Note that, if a vehicle is at a certain location $q_0$ at a time $t_0$, the time at which $sp_i$ will be served is deterministically computed as
\[
    \hat t_i = t_0 + \sum_{j=1}^i [\tau_v(q_{j-1},q_j) + b_j +t_{l,j}]
\]
where $b_j$ is the time needed for a passenger to board/alight to/from a vehicle, depending of whether $sp_j$ is a pick-up or drop-off. The value $t_{l,j}$ is the time lost for accelerating after stop point $sp_{j-1}$ and decelerating before $sp_j$. Observe that it can happen that two consecutive stop points are in the same location, i.e., $q_{j-1} = q_j$. In this case, $\tau_v(q_{j-1},q_j)=0$ and $t_{l,i}=0$, meaning that the pick-up and drop-off operations are performed one after the other, without any vehicle movements in between. For a more detailed explanation of the vehicle movement model and the computation of the time lost in accelerating and decelerating, refer to \S~3.3 of~\cite{DiMaria2018}.
Each vehicle has a capacity $C$. When constructing schedules, we must ensure that the number of passengers does not exceed $C$. Suppose a vehicle has at a certain point in time $n_0$ passengers on board and a certain schedule $S_v = [sp_1,sp2,\dots]$. Obviously, the number of passengers changes only when certain $sp_i$ is served. Let us introduce 
\[
    \rho(sp_i) = 
    \begin{cases}
    +1 & \text{if }sp_i\text{ is a pickup}\\
    -1 & \text{if }sp_i\text{ is a dropoff}
    \end{cases}
\]

The number of passengers onboard after $sp_i$ has been served is:
\begin{flalign}
    \label{eq:constr2}
    n_{i,v} = n_0 + \sum_{j=1}^i \rho(sp_i) < C
\end{flalign}

This inequality must hold for any vehicle $v$ and any stop point $sp_i$ in $S_v$.

\begin{defi}
A schedule is \emph{feasible} if it respects constraints~\eqref{eq:constr1},\eqref{eq:constr2}, i.e., if all pick-ups and drop-offs are guaranteed to be served within the expressed time requirements and the vehicle capacity is never exceeded.
\end{defi}

\section{Dispatching algorithm}
\label{sec:algorithm}
We describe in this section how the system we consider associates users to available vehicles and how routes are constructed. We assume for simplicity all the logic is centralized in a single coordinator. Our dispatching logic is implemented by an on-line algorithm: every time a request arrives to the system, we associate it to a vehicle and we modify its schedule accordingly. The dispatch is based on the concept of schedule \emph{cost}. 

\begin{defi}
Consider any vehicle $v$ at a certain time instant, its schedule $S_v = [sp_1, sp_2, \dots, sp_n]$ and its current location $q_0$. Following the notation of \S~\ref{sec:schedules}, the cost of that schedule is the amount of time needed to execute it completely, i.e.
\begin{flalign}
    \label{eq:cost}
    c(S_v, q_0) = \sum_{j=1}^n [\tau_v(q_{j-1},q_j) + b_j +t_{l,j}]
\end{flalign}
\end{defi}

The dispatching algorithm is given in Alg.~\ref{alg:dispatching-algorithm}. Let us consider a user with origin $o$, destination $d$, time constraint $\Delta t$ and suppose the user ``appeared'' at time $t$ (see \S~\ref{sec:requests}). Suppose the user arrives at the ingress stop at time $t_1\ge t$ and generate her request from there. The corresponding pair of stop points are $sp^a=\left(\phi(o),t_1,\Delta t,\text{pickup} \right)$ and $sp^b = \left( \phi(d), t_2, \Delta t, \text{dropoff} \right)$, as in \S~\ref{sec:requests}.

The dispatching algorithm seeks the ``best vehicle'' to serve this request.
In order to do so, we associate a cost $c(v)$ to each vehicle and we choose the vehicle with the lowest cost. To compute $c(v)$ for a vehicle $v$, we compute different \emph{tentative schedules}, based on the following explanation. We first try to find the best placement for the pick-up stop-point $sp^a$. To this aim, we create tentative schedules $S_v^i$ for $i=1,\dots,n$, obtained placing $sp^a$ in position $i$:
\[
    S_v^i = [sp_1,\dots,sp_{i-1},sp^a,sp_{i+1},\dots,sp_n]
\]

If $S_v^i$ is feasible, we compute its cost $c(S_v^i,q_0)$. If for all $i=1,\dots,n$ we have that $S_v^i$ is infeasible, it means the request cannot be associated to the vehicle. We indicate this by assigning $c(v)=\infty$. Otherwise, we take the best tentative schedule, i.e., we take the $i^*$ such that $S_v^{i*}$ has minimum cost. We than try to place the drop-off stop point $sp^b$ into the tentative schedule $S_v^{i*}$. In order to do so, we consider different other tentative schedules $S_v^{i*,j}$, where $sp^b$ is placed in the $j$-position of $S_v^{i*}$, with $j>i$:
\[
    S_v^{i*,j} = [sp_1,..,sp_{i^*-1},sp^a,sp_{i^*+1},..,sp_{j-1},sp^b,sp_{j+1},..,sp_n]
\]

If for all $j>i$, we have that $S_v^{i*,j}$ is infeasible, it means that the request cannot be served by vehicle $v$, which we denote with $c(v)=\infty$. Otherwise, we take $j^*$ such that the cost $c(S_v^{i*,j*},q_0)$ is the minimum among $c(S_v^{i*,j},q_0)$ for all $j>i$. We set $c(v) = c(S_v^{i*,j*},q_0)$.

The operations described above allow to compute, for each arriving request, (i) the cost $c(v)$ of serving that request with any vehicle $v$ and, if $c(v)<\infty$ (ii) the modified schedule $S_v^{i*,j*}$ including the pick-up and the drop-off related to that request. The best vehicle is the one that guarantees the lowest cost, i.e. $v^* = \arg\min_v c(v)$. The request is thus assigned to $v^*$ and the schedule $S_v$ is modified in $S_v^{i*,j*}$. Note that if for all vehicles $v$ we have $c(v)=\infty$, it means there is no vehicle which can satisfy that request and we discard it.

We repeat the process every time a new request arrives. Note that the dispatching algorithm continuously modify vehicles route in order to satisfy incoming requests with the lowest cost possible, while at the same time meeting the requirements of the incoming request as well as the requirements of requests that have been previously assigned.

\begin{algorithm}[ht]
\caption{Dispatching Algorithm (running for any incoming request)}
\label{alg:dispatching-algorithm}
\begin{small}
\begin{algorithmic}[1]
\renewcommand{\algorithmicrequire}{\textbf{Input:}}
\renewcommand{\algorithmicensure}{\textbf{Output:}}
\REQUIRE Request stop-points $sp^a, sp^b$. Current vehicle location $q_0$ and current vehicle schedule $S_v$ of any vehicle $v$.
\ENSURE  The vehicle $v^*$ that will serve the request. Its modified schedule $S_{v*}^{i*,j*}$.
%
%\STATE $S = \emptyset$ Final computed schedules ordered by Cost
\FOR {each vehicle $v$}
  \STATE Consider the current schedule $S_v=[sp_1,\dots,sp_n]$.
  \STATE $c(v) = \infty$; $S^*_v = S_v$; $i^* := 1$% $S^{i*}_v = S^*_v$
  \item[]
  \item[]// Find the best placement for the pick-up stop point.
  \FOR{$i \; in \; {1, ..., n}$}
    \STATE $S_v^i := [sp_1,\dots,sp_{i-1},sp^a,sp_{i+1},\dots,sp_n]$
    \IF{$S_v^i$ is feasible}
        \STATE Compute $c(S_v^i, q_0)$ as in~\eqref{eq:cost}.
        \IF{$c(S_v^i, q_0) < c(v)$}
            \STATE $c(v) := c(S_v^i, q_0)$
            \STATE $i^* := i$
        \ENDIF
    \ENDIF
  \ENDFOR
  \STATE \IF{$c(v) < \infty$}
  \item[]
            // Find the best placement for the drop-off.
            %\State $sp^a_{cost} = c(v)$
            %\State $S^{i*}_v = S^*_v$
            \STATE $c(v) := \infty$; $j^*:=i+1$
            \FOR{$j = {i^*+1, ..., n}$}
              \STATE 
                  $S_v^{i*,j} = [sp_1,..,sp_{i^*-1},sp^a,sp_{i^*+1},..,sp_{j-1},sp^b,sp_{j+1},..,sp_n]$
              \IF{$S_v^{i*,j}$ is feasible}
                \STATE Compute $c(S_v^{i*,j}, q_0)$ as in~\eqref{eq:cost}.
        
                \IF{$c(S_v^{i*,j}, q_0) < c(v)$}
                    \STATE $c(v) := c(S_v^{i*,j}, q_0)$
                    \STATE $j^* := j$
                \ENDIF

                     \ENDIF
            \ENDFOR
         \ENDIF
   \ENDFOR
    
\STATE $v^* := \arg\min_v c(v)$
\end{algorithmic}
\end{small}
\end{algorithm}

%%%%%%%%%%%%%%%%%%%%%%%%%%%%%%%%%
%%%%%%%%%%% RESULTS %%%%%%%%%%%%%
%%%%%%%%%%%%%%%%%%%%%%%%%%%%%%%%%
\section{Results}
We present in this section the results of our simulation campaign that we ran on AMoDSim~\cite{DiMaria2018}, an open-source simulation. The code to reproduce the results presented here is available in the AMoDSim's repository.
We will first describe the scenarios considered and then study the impact of consolidation on the operations of the system (capacity, vehicle routes, sharing degree) and on the QoS.

\subsection{Scenarios}
As in~\cite{DiMaria2018}, the network consists in a Manhattan grid of 60 Km\textsuperscript{2}, representative of the city of Manhattan with static travel times as in~\cite{Mahmassani2018}. The distance between east-west-oriented roads and north-south oriented roads is 80m and 200m, respectively.  

Requests are generated as a Poissonian process with a certain rate $r$ ranging from 20 req/h/Km2 to 320 req/h/Km2, compatible with to the scenarios in the literature~  \cite{Alonso-Mora2017,Jung2013}. 
 The origins and destinations are uniformly distributed in the surface. The time constraint $\Delta t$ (see \S~\ref{sec:requests}) is fixed to 20 minutes. Note that not all the requests will be actually generated: if the origin-destination pair of a trip is within a distance of 1.6 Km, we assume the traveler prefers to walk and does not call the system.

Vehicle cruising speed is 35 Kmph. The acceleration and deceleration of the vehicle after/before it stops is 1.676 mpss in absolute value, which gives an additional time lost for accelerating and decelerating of 11.5 s (see \S~3.3 of~\cite{DiMaria2018} and references therein for a description of the vehicle movement model). The walk speed is constant and fixed to 3.6 Kmph. The board and alight times are 5s and 10s, respectively, as in~\cite{Elpern-Waxman2017}. Observe that we do not want to make assumptions on the capacity of the vehicles, as we are exploring a service which is in the middle between a taxi-like and a bus-like service. For this reason, we set in our simulations the seat capacity to $C=45$ passengers, which in our simulation is never filled in our scenarios and thus corresponds de facto to not being limited by a pre-determined capacity. We consider a relatively small fleet (with respect to~\cite{Alonso-Mora2017}), of 500 or 1000 vehicles, since our challenge is to offer a high capacity service with relatively low cost operations. In all the simulations, if not explicitly specified, the fleet size is 1000 and the request rate 320 req/h/Km2, which we fix as default values. 
We consider different values of spacing between admitted stops, ranging from 80m, which corresponds to admit stops, thus forcing non consolidation, up to 860m.
All results are collected during 4h simulation, to wash out transient effects.

\subsection{Impact on system capacity}
\label{sec:impact-operator}

\begin{figure}[h]
\setlength{\belowcaptionskip}{-20pt}
\begin{center}
\begin{subfigure}[b]{.49\linewidth}
\includegraphics[width=\linewidth]{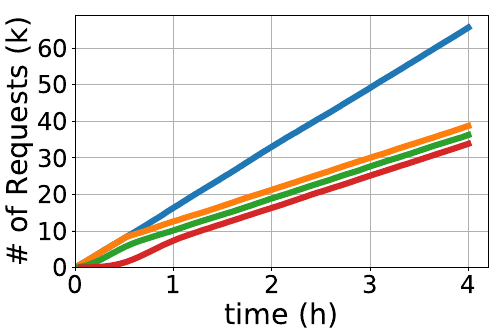}
\end{subfigure}
\begin{subfigure}[b]{.49\linewidth}
\includegraphics[width=\linewidth]{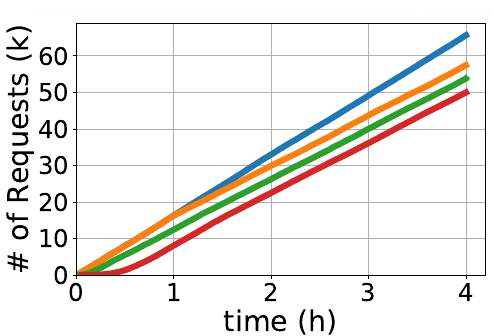} %\label{fig:perf_seats}
\end{subfigure}

\begin{subfigure}[b]{.99\linewidth}
\includegraphics[width=\linewidth]{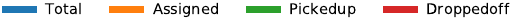}
\end{subfigure}
\caption{Load of the system with stop spacing 80m (left) and stop spacing 860m (right).}
\label{fig:load-over-time}
\end{center}
\end{figure}

We show in this section the benefits of limiting the density of admissible stops when the demand, i.e., the request rate, is high with respect to the number of vehicles. In Fig.~\ref{fig:load-over-time} we represent the number of requests over the simulation time. In particular, we represent the total number of requests sent by the users, the requests that the system has been able to assign to some vehicle, the number of drop-offs and pick-ups. Recall that a request remains unassigned if no vehicle can serve it within the time-limits. In this case, the requests are rejected, i.e., the user would receive a message saying that the system is not able to handle her request. The number of assigned requests is thus representative of the \emph{system capacity}.

First, observe the load of the system with no consolidation (Fig.~\ref{fig:load-over-time}-left). The system is much slower to assign requests then the rate of incoming requests, which denotes overload. Note that the overload is instead alleviated reducing the stop density, e.g., increasing the stop spacing (Fig.~\ref{fig:load-over-time}-right): by consolidating the demand in fewer locations, the system is able to better serve a high request rate, thus offering more capacity.

\begin{figure*}[h]
\setlength{\belowcaptionskip}{-20pt}
    \begin{center}
    \begin{subfigure}[b]{.32\linewidth}
	\includegraphics[width=\linewidth]{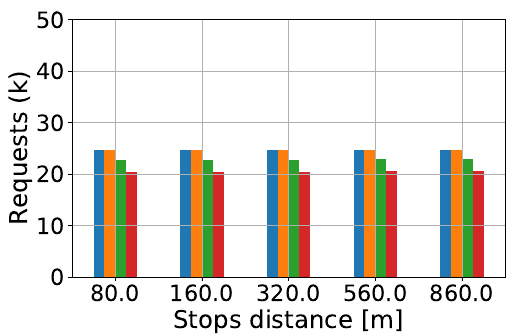}
	\end{subfigure}
    \begin{subfigure}[b]{.32\linewidth}
	\includegraphics[width=\linewidth]{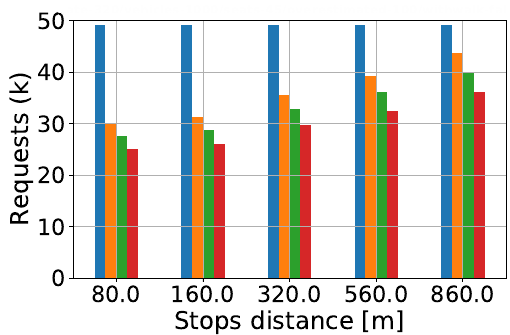}
	\end{subfigure}
    \begin{subfigure}[b]{.32\linewidth}
	\includegraphics[width=\linewidth]{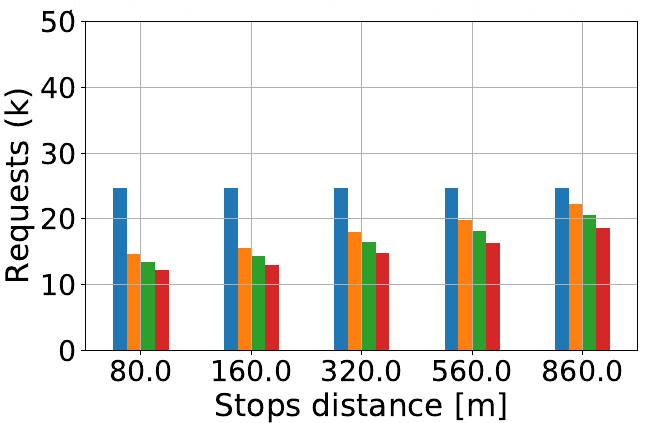}
	\end{subfigure}

    \begin{subfigure}[b]{.49\linewidth}
	\includegraphics[width=\linewidth]{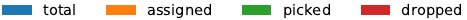}
	\end{subfigure}
	
	\caption{Load of the system after 3h with 160 req/h/Km2 and 1000 vehicles (left), 320 req/h/Km2 and 1000 vehicles (center) and 160 req/h/Km2 and 500 vehicles.}
    \label{fig:load-3h}
    \end{center}
\end{figure*}

We validate this finding by depicting the number of requests after 3h, with different values of stop spacing (Fig.~\ref{fig:load-3h}). We chose 3h, since it is usually considered the maximum duration of peak periods in transportation.

We observe in the plot in the center, where 1000 vehicles are deployed and the request rate is 320 req/h/Km2, that the responsiveness of the system in terms of requests assigned, picked-up and dropped-off consistently improves when reducing stop density. Note that this holds only in situations where the request rate is high with respect to the deployed vehicles, i.e., in the central plot and also in the plot on the right, where 160 req/h/Km2 are served by just 500 vehicles. If we instead maintain the same request rate 160 req/h/Km2, but we serve it with more vehicles (figure on the left), we see consolidation is not important, as the system is not overloaded anyways. We obtain the same picture, which we omit here, when further reducing the demand. Therefore, the correct stop spacing to avoid overload depends on the proportion between request rate and the fleet size and consolidation by reducing stop density becomes important only when this proportion is high.

%%%% ROUTE EFF
\subsection{Impact on route efficiency}
\label{sec:route-efficiency}
\begin{figure}[h]
  \centering
	\includegraphics[scale=0.30]{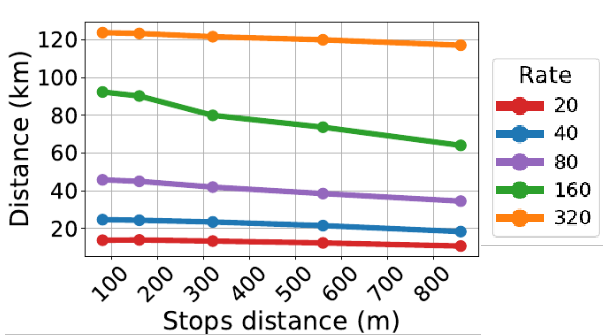}
	\caption{Mean distance traveled by a vehicle}
    \label{fig:km-traveled}
\end{figure}

Fig.~\ref{fig:km-traveled} shows that aggregating demand in fewer stops reduces the kilometers traveled by the vehicles and thus operational cost. To explain this, we study how ``efficient'' are the routes constructed by the dispatching algorithm.

\begin{figure}[h]
\setlength{\belowcaptionskip}{-20pt}
\begin{center}
\begin{subfigure}[b]{.49\linewidth}
\includegraphics[width=\linewidth]{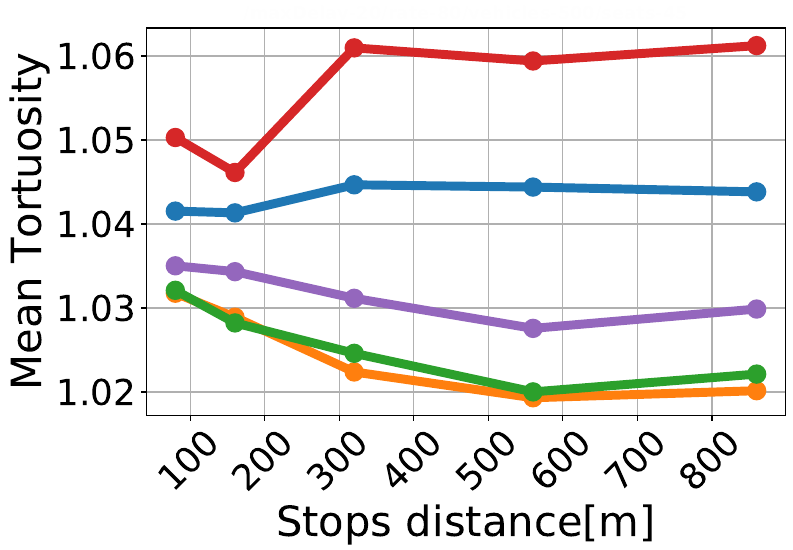}
\end{subfigure}
\begin{subfigure}[b]{.49\linewidth}
\includegraphics[width=\linewidth]{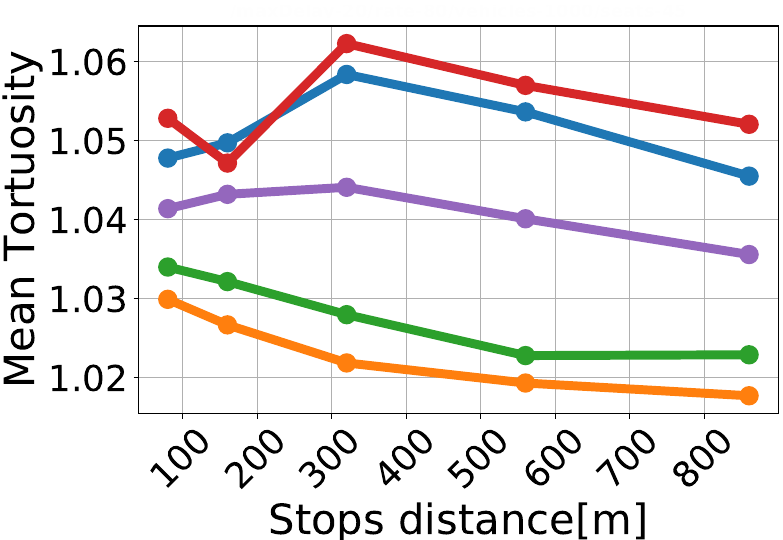} %\label{fig:perf_seats}
\end{subfigure}

\begin{subfigure}[b]{.75\linewidth}
\includegraphics[width=\linewidth]{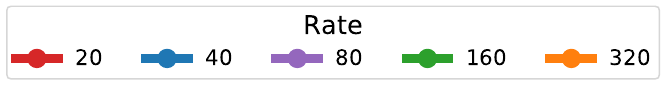}
\end{subfigure}
\caption{Mean tortuosity of a vehicle route with a fleet size of 500 vehicles (left) or 1000 vehicles (right).}
\label{fig:tortuosity}
\end{center}
\end{figure}

We adopt the convention that a route is  `efficient' if it visits all the stop locations contained in it traveling a small distance. The more routes are tortuous, e.g., passing more times than necessary through the same stops or making un-necessary long detours, the less the system is efficient. To quantify this aspect we define a measure of \emph{tortuosity}.

Let us consider a trajectory of a vehicle $v$, i.e., the sequence $q_{1},\dots,q_{n}$, where $q_i$ is the location of stop point $sp_i$. Let us truncate the trajectory starting
from $i$ and ending in $i+H$, i.e., thus considering the sub-trajectory
$q_{i},\dots,q_{i+H}$. We call the parameter $H\in\mathbb{N}$ \emph{horizon}.
Let us consider the length $\ell(q_{i},\dots,q_{i+H})$ of this subtrajectory.
Starting from $q_{i}$, we can visit all the other points $\{q_{i+1},\dots,q_{i+H}\}$
in $H!$ ways, each associated with a length. Let $m_{q_{i}}\{q_{i+1},\dots,q_{i+H}\}$
the minimum length to visit all the other points starting from $q_{i}$.
We define the tortuosity of the trajectory at the $i$-th stop-point
as
\[
T(i,v)\equiv\frac{\ell(q_{i},\dots,q_{i+H})}{m_{q_{i}}\{q_{i+1},\dots,q_{i+H}\}}
\]

The tortuosity of a vehicle is the average of the tortuosity at all
its stop-points:
\[
T(v)=\frac{1}{n-H}\sum_{i=1}^{n-H}T(i,v)
\]

The smaller the tortuosity, the more efficient the route of the vehicle.
Note that the best we can have is $T(v)=1$ meaning that at every stop
point the vehicle chooses the shortest trajectory. Note also that the shortest trajectory may be impossible for a vehicle to follow, if it violates the time constraints of \S~\ref{sec:schedules}. Therefore, one must not be tempted to associate a physical meaning to the measures of tortuosity in absolute. They are just useful when comparing different system configurations, which we do in this section. The results we report are the tortuosity, averaged across als vehicles, for a horizon $H=4$.

We observe in Fig.~\ref{fig:tortuosity} that when the rate of requests is high consolidating demand around fewer stops helps decreasing the tortuosity, as expected. This explain the reduction in kilometers traveled (Fig.~\ref{fig:km-traveled}). However, the opposite effect is recorded with low request rates. One hypothesis to explain this behavior at low rates is that, while serving the users on board, each vehicle does not collect enough requests in a unit of time so that it can optimize the route. However, this deserves further investigation in future work.

It is interesting to emphasize again that all results do not depend on just stop spacing but also on the fleet size. Fig.~\ref{fig:tortuosity} shows the beneficial impact of consolidation shows off earlier (i.e., already with smaller request rates) with a fleet of 500 vehicles than 1000 vehicles.

\subsection{Impact on sharing degree}

\begin{figure}[h]
\setlength{\belowcaptionskip}{-10pt}
  \centering
	\includegraphics[scale=0.2]{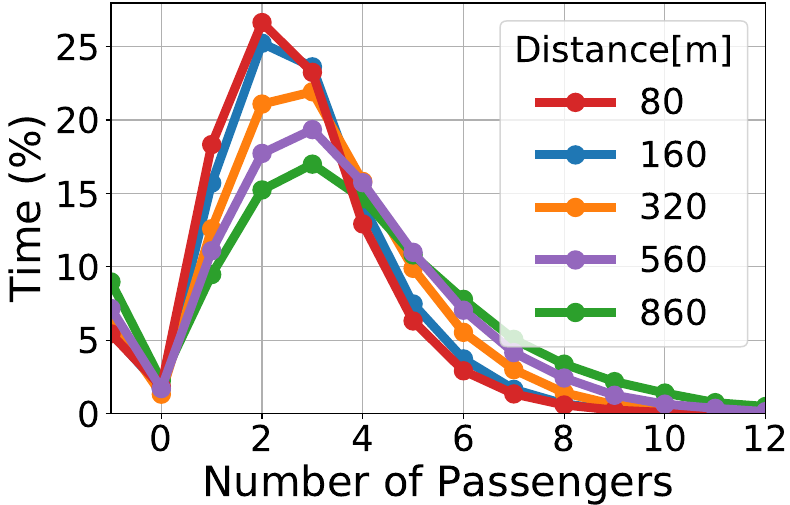}
	\caption{Vehicle occupation: percentage of time each vehicle spends with different numbers of passengers onboard. The x-value -1 denotes that the vehicle is idle.}
    \label{fig:sharing-degree}
\end{figure}

Conceptually, the \emph{sharing degree} of the system is its ability to aggregate together multiple trips in the same vehicle at the same time. We quantify the sharing degree as the average percentage of time a vehicle spends with certain number of passengers on-board. Fig.~\ref{fig:sharing-degree} shows that consolidating demand around fewer stops helps increasing the sharing degree: each vehicle is able to aggregate more trips. This explains why, with a fixed amount of vehicles, the system offers more capacity to serve high volume of requests, when more consolidation is put in place (less stops), as discussed in \S~\ref{sec:impact-operator}. Observe that with consolidation vehicles behave for a non-negligible amount of time as minibuses.

%%%%% QOS
\subsection{Impact on quality of service}

\begin{figure}[h]
\setlength{\belowcaptionskip}{-10pt}
  \centering
	\includegraphics[width=0.4\linewidth]{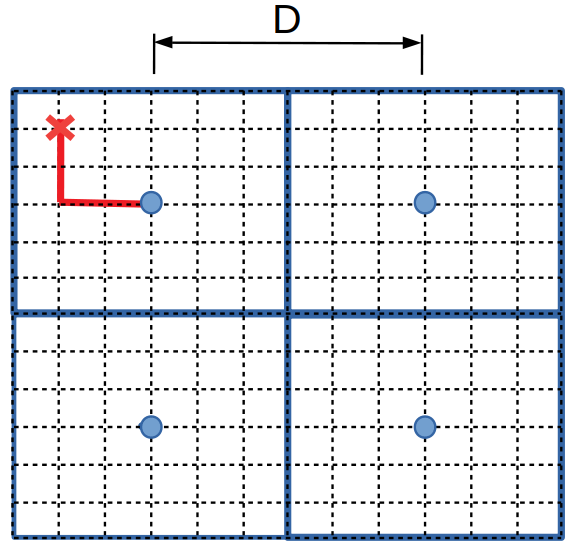}
	\caption{Computation of the mean ingress time. Dotted lines indicate the road network. The 4 dots represent 4 admitted stops and the cross the origin of a user. The surface is divided in 4 Voronoi cells. The L-shaped line is the y-x route followed by the user to walk from her origin to the closest stop location. }
    \label{fig:ingress-egress}
\end{figure}

While in \S~\ref{sec:impact-operator} we have shown that it is advantageous for an operator to limit the density of possible stop locations, we need to study possible negative side effects on the quality of service offered to the users. It is obvious to observe that the more the stop spacing, the longer the ingress and egress times suffered by users.
It is possible to compute the mean ingress and egress time for any given value of stop spacing with the help of Fig.~\ref{fig:ingress-egress}.

Let us focus on the mean ingress distance walked by a user (the mean egress distance would be computed similarly). Let us denote with $D$ the stop spacing. Since we assume that a user always goes to the closest stop, we can partition the entire studied surface in a Voronoi tassellation where the seeds correspond to the stops. A Voronoi cell of a stop is the set of points that are closer to that stop than to any other. Since our stops are by construction located in a regular grid, the Voronoi cells are simply squares with edge measuring $D$. Let us consider any origin point $(x,y)$ in any certain Voronoi cell around a stop $(x_0,y_0)$. The distance traveled by a user to go from that origin to the ingress stop is $z=|x-x_0|+|y-y_0|$. Since $x$ and $y$ are generated uniformly at random, we are sure that $|x-x_0|\simeq\mathcal{U}(0,D/2)$. The same holds for $|y-y_0|$. Therefore the mean walked distance is~$\mathbb{E}[z]=D/4 + D/4 = D/2$. Since we assume a fixed walking speed $v_w = 3.6$~Km/h, we can immediately compute the mean ingress and egress time based on the value $\mathbb{E}[z]$. With maximum consolidation, i.e., $D=860$m, the mean ingress time is $D/2 / v_w = 7$ min and the maximum value of ingress time is $14$ min.
In other words, the ingress and egress time imposed on users to aggregate the demand around fewer stops is, in our scenario, in a limited range of acceptable values.

\begin{figure}[h]
\setlength{\belowcaptionskip}{-20pt}
\begin{center}
\begin{subfigure}[b]{.49\linewidth}
\includegraphics[width=\linewidth]{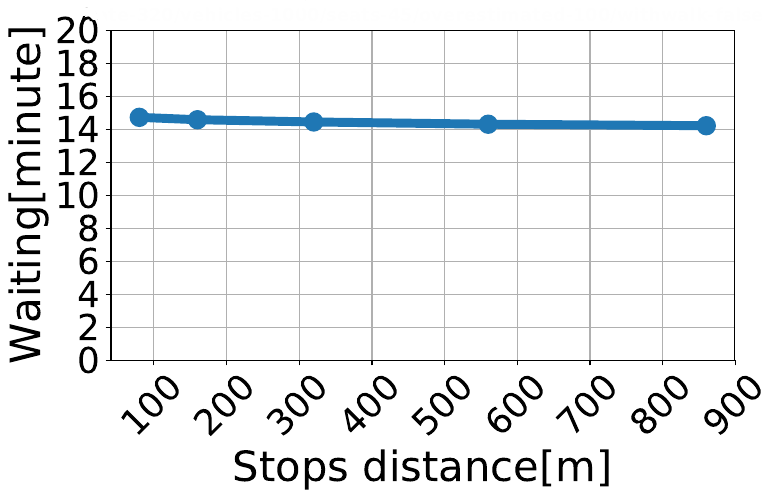}
\end{subfigure}
\begin{subfigure}[b]{.49\linewidth}
\includegraphics[width=\linewidth]{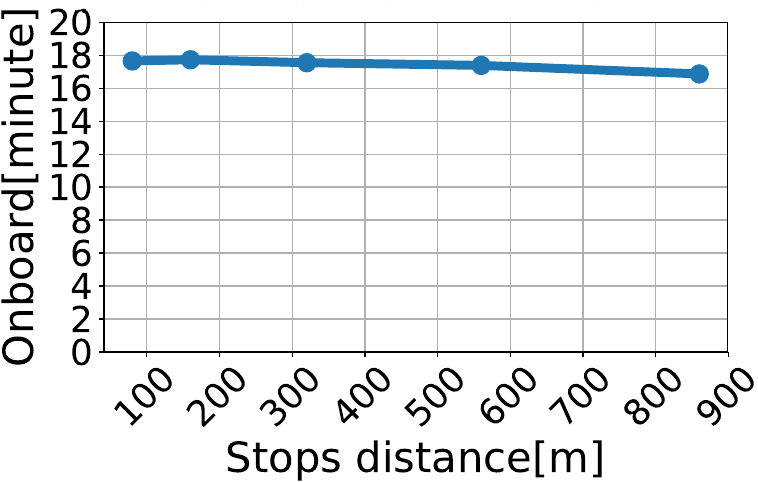} 
\end{subfigure}
\caption{Waiting time and onboard time.}
\label{fig:qos}
\end{center}
\end{figure}

\begin{figure}[h]
\setlength{\belowcaptionskip}{-20pt}
\begin{center}
\begin{subfigure}[b]{.42\linewidth}
\includegraphics[width=\linewidth]{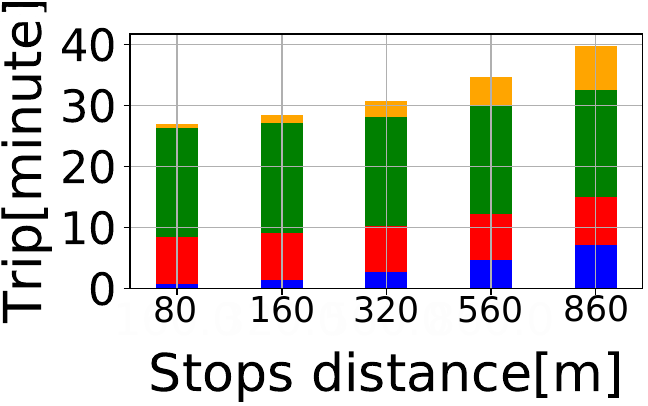}
\end{subfigure}
\begin{subfigure}[b]{.54\linewidth}
\includegraphics[width=\linewidth]{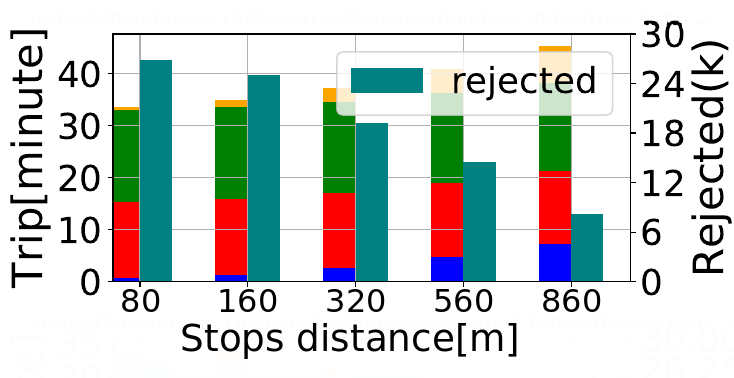} 
\end{subfigure}

\begin{subfigure}[b]{.99\linewidth}
\includegraphics[width=\linewidth]{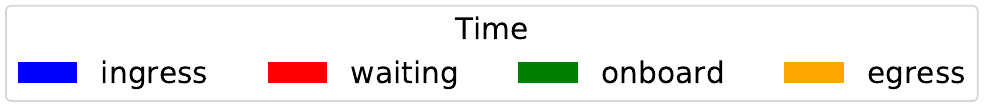}
\end{subfigure}

\caption{Total travel time with 20 req/h/Km2 (left) and 320 req/h/Km2. On the left, no requests are rejected.}
\label{fig:total-travel-time}
\end{center}
\end{figure}

It is interesting to observe (Fig.~\ref{fig:qos}) that the waiting time slightly improves with less stop density and the on-board time (time spent by the user on board of a vehicle) is practically not affected, thanks to the increased overall efficiency of routes (\S~\ref{sec:route-efficiency}) that finally benefits users as well. However, we should recall that the more stop spacing, the more ingress and egress time suffered by users. It is thus important to compute the total travel time, which is the time passed from the moment the user appears at her origin to the moment where she arrives at the intended destination. The total travel time is the sum of ingress, waiting, onboard and egress times. The total travel time is plotted in Fig.~\ref{fig:total-travel-time}, which shows, as expected, that maximum consolidation (stop spacing 860m) increases total travel time. This increase is severe with a low request rate (33\% of increase with $r=20$ req/h/Km2), while it is less important with high request rate (23\% with $r=320$ req/h/Km2). This confirms that consolidation is deleterious with low demand, since it reduces QoS without improving capacity, while it is desirable when high demand has to be satisfied, as high capacity can be achieved in exchange for an acceptable loss in QoS. Fig.~\ref{fig:total-travel-time} helps also to visualize the trade-off this paper is about: the more the consolidation, the higher service capacity (less requests are rejected) but the lowest the QoS for the requests that manage to be served.

\section{Negative results and future extensions}
%When preparing a scientific paper, authors often try different system settings but when ``packing'' the final version of the paper only report the results of the best of their configurations. We think instead that it is useful for the scientific community to also share negative results related to configurations that did not work out and other configurations that have been considered but that have not been tested and are potentially promising. This is the goal of this section.

Observe that the systems model we adopt is one of the many possible. A first element we emphasize is that the time constraint $\Delta t$ does not include the walking time. For example, if a user specifies a certain $\Delta t$, it means she wants to wait at most $\Delta t$ after she arrives at $\phi(o)$, independent of whether her ingress time is high or small. Moreover, the time constraint $\Delta t$ also means that if she arrives at the stop at time $t_1$ and the minimum time to drive from $\phi(o)$ to $\phi(d)$ is $\tau_v \left(\phi(o),\phi(d) \right)$, she wants to arrive at $\phi(d)$ within a time $t_1 + \tau_v\left( \phi(o),\phi(d) \right) + \Delta t$, independent of whether the ingress and egress time are high or small. 
%
%In this work, we chose to adopt these simplifying assumptions, which penalize Quality of Service (QoS), and check a-posteriori that QoS is anyways not excessively degraded (see numerical results).
%
We have also experimented with alternative formulations, which take into account the ingress and egress time in the expression of time constraints, in order, for example, to compensate users with higher ingress time by making them wait less. However, we observed that the results were worse both in terms of operational costs for the operator and user QoS. We noticed that by including ingress and egress time in the time constraints translates in too strict conditions for the dispatching algorithm to construct efficient schedules. For instance, if a user has a $\Delta t=20$ min and has to walk 14 minutes, the system would see the request after the user arrives at the stop and would have to satisfy it within 6 minutes waiting time, which leaves no sufficient degree of freedom for the construction of vehicle routes: if a vehicle is able to make a detour to satisfy this strict constraint it would do it, even if it could be highly inefficient. While this type of behavior would be acceptable in a taxi-like service with few requests with respect to the fleet size, we observed that it is not appropriate for high capacity MoD, which should instead resemble a bus-like system, i.e., should be less sensitive to the single user time requirements and more sensitive to the overall system performance. 

Observe that the system we presented can re-configure its behavior just by changing the admitted stop density and we showed that it has a relevant impact on capacity and QoS, although the dispatching algorithm does not change at all. We will explore in the future other algorithmic strategies which can improve both QoS and capacity by changing the algorithm behavior based on the value of stop density or input demand.

Another margin of improvement is given by the introduction of a reservation. While in the model adopted now a user sends a request only after she arrives at the closest stop, it would be possible to let her send the request immediately after she ``appears'' at the origin, specifying that she will arrive at the closest stop in a certain time interval. The algorithm could thus be able to handle this request in advance. We expect improvement in both operational cost metrics and QoS and we plan to study such possibility as part of future work.

To make our study more realistic, we also plan to introduce variable speed for vehicles, depending on the conditions of traffic, and to let traffic be impacted by MoD vehicles. However, as the goal of this paper is to study the high level impact of demand consolidation in MoD, the lack of these elements of realism does not invalidate the findings presented here.

%\begin{figure}[htbp]
%\centerline{\includegraphics{fig1.png}}
%\caption{Example of a figure caption.}
%\label{fig}
%\end{figure}

\section{Conclusion}
In this paper we studied the importance of consolidation in Mobility on Demand (MoD) systems. We observed that by varying the parameter determining the spacial density of admitted stop locations, we can favor high capacity or Quality of Service (QoS). We showed that the benefits of consolidation cannot be evaluated in absolute. They depend on the vehicle fleet size and the amount of demand (request rate). In particular, consolidation is beneficial when the demand is high with respect to the number of available vehicles, but is deleterious otherwise.

Ride-sharing systems have mostly no-consolidation at all, being mainly door-to-door services. On the opposite side, fixed-route traditional transit has maximum consolidation, as only a restricted amount of stops are served. We show instead that a system can be engineered so as to function in any intermediate operational mode between those two extremes, varying the level of consolidation. This paper is a first step toward an investigation of such systems, which we will pursue in our future work.

%\section{Acknowledgement}
%This work is supported by the CLARA - CLoud plAtform and smart underground imaging for natural Risk %Assessment - project, funded by the Italian Ministry of Education, Universities and Research, within %the “Smart Cities and Communities and Social Innovation” initiative.

\bibliographystyle{IEEEtran}
\bibliography{On_the_Importance_of_demand_Consolidation_in_Mobility_on_Demand}

\begin{thebibliography}{10}
\providecommand{\url}[1]{#1}
\csname url@samestyle\endcsname
\providecommand{\newblock}{\relax}
\providecommand{\bibinfo}[2]{#2}
\providecommand{\BIBentrySTDinterwordspacing}{\spaceskip=0pt\relax}
\providecommand{\BIBentryALTinterwordstretchfactor}{4}
\providecommand{\BIBentryALTinterwordspacing}{\spaceskip=\fontdimen2\font plus
\BIBentryALTinterwordstretchfactor\fontdimen3\font minus
  \fontdimen4\font\relax}
\providecommand{\BIBforeignlanguage}[2]{{%
\expandafter\ifx\csname l@#1\endcsname\relax
\typeout{** WARNING: IEEEtran.bst: No hyphenation pattern has been}%
\typeout{** loaded for the language `#1'. Using the pattern for}%
\typeout{** the default language instead.}%
\else
\language=\csname l@#1\endcsname
\fi
#2}}
\providecommand{\BIBdecl}{\relax}
\BIBdecl

\bibitem{Lo2018}
J.~Lo and S.~Morseman, ``{The Perfect uberPOOL: A Case Study on Trade-Offs},''
  in \emph{Ethnographic Praxis in Industry Conf.}, 2018.

\bibitem{Hawkins2018}
A.~J. Hawkins, ``{Uber Express Pool Offers the Cheapest Fares Yet in Exchange
  for a Little Walking},'' 2018.

\bibitem{Quadrifoglio2009}
L.~Quadrifoglio and X.~Li, ``{A methodology to derive the critical demand
  density for designing and operating feeder transit services},''
  \emph{Transportation Research Part B}, vol.~43, no.~10, pp. 922--935, 2009.

\bibitem{Guo2018}
Q.~W. Guo, J.~Y. Chow, and P.~Schonfeld, ``{Stochastic dynamic switching in
  fixed and flexible transit services as market entry-exit real options},''
  \emph{Transportation Research Part C}, vol.~94, pp. 288--306, 2018.

\bibitem{Koh2018}
K.~Koh, C.~Ng, and D.~Pan, ``{Dynamic Bus Routing : A study on the viability of
  on-demand high-capacity ridesharing as an alternative to fixed-route buses in
  Singapore},'' \emph{IEEE ITSC}, 2018.

\bibitem{Chandra2013}
S.~Chandra and L.~Quadrifoglio, ``{A model for estimating the optimal cycle
  length of demand responsive feeder transit services},'' \emph{Transportation
  Research Part B: Methodological}, vol.~51, pp. 1--16, 2013.

\bibitem{DiMaria2018}
A.~{Di Maria}, A.~Araldo, G.~Morana, and A.~{Di Stefano}, ``{AMoDSim: An
  Efficient and Modular Simulation Framework for Autonomous Mobility on
  Demand},'' in \emph{Internet of Vehicles Conference}, 2018.

\bibitem{Basu2018b}
R.~Basu, A.~Araldo, and {et Al.}, \emph{{Automated Mobility-on-Demand vs. Mass
  Transit: A Multi-Modal Activity-Driven Agent-Based Simulation Approach}},
  2018.

\bibitem{Berbeglia2010a}
G.~Berbeglia, J.-F. Cordeau, and G.~Laporte, ``{A Hybrid Tabu Search and
  Constraint Programming Algorithm for the Dynamic Dial-a-Ride Problem},''
  \emph{INFORMS Journal on Computing}, vol.~24, no.~3, pp. 343--355, 2012.

\bibitem{Nunez2013}
A.~A. N{\'{u}}{\~{n}}ez, D.~A. S{\'{a}}ez, and C.~E. Cort{\'{e}}s,
  \emph{{Hybrid Predictive Control for Dynamic Transport Problems}}, 2013.

\bibitem{Winter2016a}
K.~Winter, O.~Cats, G.~H. d.~A. Correia, and B.~van Arem, ``{Designing an
  Automated Demand-Responsive Transport System: Fleet Size and Performance
  Analysis for a Campus–Train Station Service},'' \emph{Transportation
  Research Record}, vol. 2542, no.~1, pp. 75--83, 2016.

\bibitem{Alonso-Mora2017}
J.~Alonso-mora, S.~Samaranayake, A.~Wallar, E.~Frazzoli, and D.~Rus,
  ``{On-demand high-capacity ride-sharing via dynamic trip-vehicle
  assignment},'' \emph{Proceedings of the National Academy of Sciences of the
  United States of America}, vol. 114, no.~3, pp. 462--467, 2017.

\bibitem{Watel2018}
D.~Watel and A.~Faye, ``{Taxi-sharing: Parameterized complexity and
  approximability of the dial-a-ride problem with money as an incentive},''
  \emph{Theoretical Computer Science}, vol. 745, pp. 202--223, 2018.

\bibitem{Bosch2018}
P.~M. B{\"{o}}sch, F.~Becker, H.~Becker, and K.~W. Axhausen, ``{Cost-based
  analysis of autonomous mobility services},'' \emph{Transportation Policy},
  vol.~64, pp. 76--91, 2018.

\bibitem{Stein1978}
D.~M. Stein, ``{Scheduling Dial-a-Ride Transportation Systems},''
  \emph{Transportation Science}, vol.~12, no.~3, pp. 232--249, 1978.

\bibitem{Stiglic2015}
M.~Stiglic, N.~Agatz, M.~Savelsbergh, and M.~Gradisar, ``{The benefits of
  meeting points in ride-sharing systems},'' \emph{Trans. Res. Part B}, 2015.

\bibitem{Berbeglia2010}
G.~Berbeglia, J.~F. Cordeau, and G.~Laporte, ``{Dynamic pickup and delivery
  problems},'' \emph{European Journal of Operational Research}, vol. 202,
  no.~1, pp. 8--15, 2010.

\bibitem{EI-Geneidy2007}
A.~EI-Geneidy, J.~Strathman, T.~Kimpel, and D.~Crout, ``{Effects of Bus Stop
  Consolidation on Passenger Activity and Transit Operations},'' in
  \emph{Transportation Research Board}, 2005.

\bibitem{Li2018a}
Y.~Li, G.~Liu, Z.-L. Zhang, J.~Luo, and F.~Zhang, ``{CityLines: Designing
  Hybrid Hub-and-Spoke Transit System with Urban Big Data},'' \emph{IEEE
  Transactions on Big Data}, pp. 1--1, 2018.

\bibitem{Mahmassani2018}
M.~Hyland and H.~Mahmassani, ``{Dynamic Autonomous Vehicle Fleet Operations:
  Optimization-Based Strategies to Assign AVs to Immediate Traveler Demand
  Requests},'' \emph{Trans. Res. Part C}, vol.~92, 2018.

\bibitem{Jung2013}
J.~Jung, R.~Jayakrishnan, and J.~Y. Park, ``{Design and Modeling of Real-time
  Shared-Taxi Dispatch Algorithms},'' in \emph{TRB Annual Meeting}, 2013.

\bibitem{Elpern-Waxman2017}
J.~Elpern-Waxman, ``{Transportation Terms: Dwell Time},'' 2017.

\end{thebibliography}

%\begin{thebibliography}{00}
%\bibitem{b1} G. Eason, B. Noble, and I. N. Sneddon, ``On certain integrals of Lipschitz-Hankel type involving products of Bessel %functions,'' Phil. Trans. Roy. Soc. London, vol. A247, pp. 529--551, April 1955.
%\bibitem{b2} J. Clerk Maxwell, A Treatise on Electricity and Magnetism, 3rd ed., vol. 2. Oxford: Clarendon, 1892, pp.68--73.
%\bibitem{b3} I. S. Jacobs and C. P. Bean, ``Fine particles, thin films and exchange anisotropy,'' in Magnetism, vol. III, G. T. %Rado and H. Suhl, Eds. New York: Academic, 1963, pp. 271--350.
%\bibitem{b4} K. Elissa, ``Title of paper if known,'' unpublished.
%\bibitem{b5} R. Nicole, ``Title of paper with only first word capitalized,'' J. Name Stand. Abbrev., in press.
%\bibitem{b6} Y. Yorozu, M. Hirano, K. Oka, and Y. Tagawa, ``Electron spectroscopy studies on magneto-optical media and plastic %substrate interface,'' IEEE Transl. J. Magn. Japan, vol. 2, pp. 740--741, August 1987 [Digests 9th Annual Conf. Magnetics Japan, %p. 301, 1982].
%\bibitem{b7} M. Young, The Technical Writer's Handbook. Mill Valley, CA: University Science, 1989.
%\end{thebibliography}%

\vspace{12pt}

\end{document}